\documentstyle[12pt,iopconf,epsfig]{article}
\begin{document}

\title{Numerical Methods for Modeling Binary Neutron Star Systems}

\begin{small} 
\noindent 
(This manuscript will appear in the proceedings of the Second
Oak Ridge Symposium on Atomic and Nuclear Astrophysics.) 
\end{small} 

\author{Alan C Calder\dag\footnote{E-mail: calder@ncsa.uiuc.edu}, 
F Douglas Swesty\dag\ddag\ and Edward  Y M Wang\dag\S}

\affil{\dag\ National Center for Supercomputing Applications, 
University of Illinois,
Urbana, IL 61801, USA}

\affil{\ddag\ Department of Astronomy, University of Illinois, 
Urbana, IL 61801, USA}

\affil{\S\ 
Department of Physics, Washington University in St. Louis,
St. Louis, MO 63130, USA
and Department of Physics and Astronomy, State University
of New York at Stony Brook,
Stony Brook, NY  11794, USA}

\beginabstract
We present initial results of our study of numerical methods
for modeling neutron star 
mergers (NSMs) with simulations that perform the full 
hydrodynamic evolution required to capture tidal 
effects, particularly in the last several orbits.
Our simulations evolve the Euler equations using a
modification of the ZEUS-2D algorithm (Stone and Norman 1992).
We describe some of the difficulties
of modeling NSMs and our approaches to these difficulties,
and we discuss the motivation for the
choice of performing simulations in a co-rotating reference frame.
Our results establish what effects the choices of gravity
coupling and 
reference frame have on the numerical accuracy of the simulation.
\endabstract

\section{Introduction}

\subsection{Scientific motivation}
Neutron star mergers (NSMs) are excellent astrophysical
laboratories for the study of relativistic
astrophysics, gravitational wave astronomy, and nuclear astrophysics,
and are thought to be a possible source of observed gamma ray bursts.
General relativity predicts that binary systems of compact objects, 
such as NS-NS systems, will emit energy in the form of 
gravitational radiation. This loss of energy
will lead to the in-spiral and coalescence of the binary system,
and gravitational waves produced in these events are expected to
be observed by gravitational wave detectors 
such as LIGO (Abramovici \etal 1992). 
The limited range of frequencies of such gravitational wave detectors
makes in-spiraling compact binaries the only sources of gravitational
waves expected to be observed with high accuracy (Thorne 1995).
Post-Newtonian methods are adequate for 
the prediction of waveforms for the early stage of the in-spiral, but 
the prediction of waveforms in the later stages of the merger, when 
tidal effects and neutron star structure become important, requires 
a full three-dimensional numerical simulation.

Information about the neutron star equation of state 
as well as upper bounds on 
neutron star masses may be obtained from observed gravitational 
waveforms. Additionally, the spiral arms of the coalescing neutron 
stars may be a site of {\it r}-process nucleosynthesis 
since bombardment of
nuclei by free neutrons is expected to occur in
this environment (Lattimer \etal 1977). 
NSMs are a suggested source for the mysterious
gamma-ray bursts observed in recent years.
These events are
thought to release energy on the order of their gravitational
binding energy, $\approx 10^{53}$ erg, which is comparable to
the estimated gamma-ray burst energies, $\approx 10^{51}$
(Quashnock 1996 and Rees 1997) to $10^{53}$ erg (Woods and Loeb 1994).
A popular model for bursts at cosmological distances is the
relativistic fire-ball. NSMs are likely candidates for the source
of relativistic fire-balls, but the mechanism by which the fire-ball
develops has yet to be determined.
Simulations of NSMs can test the consistency of the merger
energetics and time scales with the estimated burst energies and observed
time scales. 

\subsection{Newtonian hydrodynamics and gravitational waveform calculation}
The simulations presented in this work are purely Newtonian,
though we have performed additional simulations that include a radiation
reaction (Wang \etal 1998). 
We assume that the neutron stars can be described as a 
compressible Newtonian fluid.
The equations of Newtonian hydrodynamics in a reference 
frame rotating at $\mbox{\boldmath $\omega$}$ are
\vspace{0.1in}
\begin{equation} \label{eqn1}
\frac{\partial \rho}{\partial t} + \nabla \cdot ( \rho {\mathbf v} ) \: = \: 0 
\end{equation}
\begin{equation} \label{eqn2}
\frac{\partial e \rho}{\partial t} +
\nabla \cdot ( e \rho {\mathbf v} ) \: = \: -(P+Q) \nabla \cdot {\mathbf v}
\end{equation}
\begin{equation} \label{eqn3}
\frac{\partial v_i \rho}{\partial t} +
\nabla \cdot ( v_i \rho {\mathbf v} ) \: = \: - \nabla (P+Q) - 
\rho \nabla \Phi -2\rho\,\mbox{\boldmath $\omega$}{\mathbf \times}{\mathbf v}
- \rho\,\mbox{\boldmath $\omega$}{\mathbf \times}\left(
\mbox{\boldmath $\omega$}{\mathbf \times}
{\mathbf r}\right)
\end{equation}
where ${\mathbf r}$ is the position, $\rho$ is the density, 
$\mathbf{v}$ is the velocity of the fluid, $e$ is
the specific internal energy, and $P$ is the pressure. $Q$ is an 
artificial viscous
stress to model the sub-resolution micro-physics occurring 
across shock fronts. The
potential $\Phi$ is the Newtonian gravitational potential.
The equations are discretized
onto a staggered Eulerian mesh by a method similar to that employed in the 
ZEUS codes (Stone and Norman 1992). 
The set of equations is closed by an equation of state, which
for this work is that of an ideal gas,
\begin{equation}
P \: = \: (\Gamma -1)E \: ,
\end{equation}
where $E = \rho e$ is the energy density and $\Gamma$ is the adiabatic
index.  Future simulations will make use of the realistic equation
of state of Lattimer and Swesty (Lattimer \etal 1985 and Lattimer and
Swesty 1990).
The Euler equations, equations 
(\ref{eqn1})--(\ref{eqn3}), are solved in two steps, an 
advection update and
a source update. The advection step makes use of second-order van 
Leer monotonic
advection with Norman's consistent advection method (Norman 1980) to evolve
the left-hand sides of equations (\ref{eqn1})--(\ref{eqn3}). The source 
update
calculates the sources and sinks of the right-hand side of equations 
(\ref{eqn2}) and (\ref{eqn3}). The solution of
Poisson's equation 
\begin{equation} \label{poisson}
\nabla ^2 \Phi ({\mathbf r}) \: = \: 4 \pi G \rho ({\mathbf r}) 
\end{equation}
gives the gravitational potential $\Phi$. 

The gravitational waveforms are calculated
from the quadrupole approximation. 
In this approximation,
the strain amplitude of the gravitational wave radiation in
the transverse traceless gauge is
\begin{equation}
h_{lm}^{TT} = {\frac{2}{r}} \, {\frac{G}{c^4}}
{\skew6\ddot{I\mkern-6.8mu\raise0.3ex\hbox{-}}}_{lm}^{\,TT},
\label{hlm-TT}
\end{equation}
where $r$ is the distance to the source, $c$ is the speed of
light, $G$ is the gravitational constant. 
${\skew6\ddot{I\mkern-6.8mu\raise0.3ex\hbox{-}}}_{lm}^{\,TT}$
is second time derivative of the transverse, traceless part 
of the reduced quadrupole moment,
${{I\mkern-6.8mu\raise0.3ex\hbox{-}}}_{lm}$,
\begin{equation}
{{I\mkern-6.8mu\raise0.3ex\hbox{-}}}_{lm} = \int\rho \,(x_l x_m -
{\textstyle{\frac{1}{3}}}
\delta_{lm} r^2) \:d^3 r \:, 
\label{Ilm}
\end{equation}
in Cartesian coordinates.
The second time derivatives of 
${I\mkern-6.8mu\raise0.3ex\hbox{-}}_{lm}$
are calculated without numerical time differentiation
(Finn and Evans 1990 and Rasio and Shapiro 1992).
Along the z-axis, the two polarizations of $h^{TT}$ are given by
\begin{eqnarray}
h_+&=&\frac{G}{c^4}\,\frac{1}{r}( {\skew6\ddot{
{I\mkern-6.8mu\raise0.3ex\hbox{-}}}}_{xx}- {\skew6\ddot{
{I\mkern-6.8mu\raise0.3ex\hbox{-}}}}_{yy}),
\label{hplus}\\
h_\times &=&\frac{G}{c^4}\,\frac{2}{r} {\skew6\ddot{
{I\mkern-6.8mu\raise0.3ex\hbox{-}}}}_{xy}.\label{hcross}
\end{eqnarray}

\subsection{Coupling gravity to the hydrodynamics}
Our code, V3D, performs the advection step before the
source step that updates the Lagrangian terms (the terms
on the right hand side of the hydrodynamics equations).
This ordering, advection before the source update, allows
the choice of computing the right-hand side of the Poisson
equation, $4\pi G \rho$, with the density at the old time
step (time lagged), the new time step (time advanced),
or the average of the two densities (time centered).
The finite difference expressions for these choices are
\begin{eqnarray*}
\left(\nabla^2 \Phi \right)_{i+{\frac{1}{2}},j+{\frac{1}{2}},
k+{\frac{1}{2}}} & = \:\:\:\:\:\:\:\:\:\:\:\:\:\:
\: 4 \pi G \rho^n_{i+{\frac{1}{2}},j+{\frac{1}{2}},
k+{\frac{1}{2}}} & \mbox{t. l.} \\
\left(\nabla^2 \Phi \right)_{i+{\frac{1}{2}},j+{\frac{1}{2}},
k+{\frac{1}{2}}} & = \:\:\:\:\:\:\:\:\:\:\:\:\:\:
\: 4 \pi G \rho^{n+1}_{i+{\frac{1}{2}},j+{\frac{1}{2}},
k+{\frac{1}{2}}} & \mbox{t. a.} \\
\left(\nabla^2 \Phi \right)_{i+{\frac{1}{2}},j+{\frac{1}{2}},
k+{\frac{1}{2}}} & = \: 4 \pi G \left(\frac{\rho^n_{i+{\frac{1}{2}},
j+{\frac{1}{2}},
k+{\frac{1}{2}}} + \rho^{n+1}_{i+{\frac{1}{2}},j+{\frac{1}{2}},
k+{\frac{1}{2}}}}{2} \right) & \mbox{t. c.} \\
\end{eqnarray*}

\subsection{Additional method details}

Our numerical method requires that material be on all of
the grid; therefore, we include a hot, low density 
($\approx$ 1 g/cm$^3$)
atmosphere as a background. The high temperature of the 
atmosphere gives the material enough energy to avoid falling 
onto the surface of the star and forming a shock, which would 
require unacceptably small time steps to evolve. 
We find thin atmospheres (1 - 10$^3$g/cm$^3$) have no effect on 
the simulation dynamics. Early simulations 
made use of a full multi-grid W-cycle algorithm to solve (\ref{poisson}),
but recent simulations (some of which are presented below) made use
of a Fast Fourier Transform (FFT) method to solve (\ref{poisson}).
The solution of the Poisson equation requires the calculation
of the potential at the grid boundaries. A direct
summation over density to get the potential on the
boundary is computationally expensive so the calculation
is performed over a coarser grid 
with the contributions of zones containing
less than $10^{-5}M_\odot$ neglected. This method works well for
``clumpy" mass distributions ({\em e.g.} two stars), and
was tested against an entire grid summation. 

We should note that many simulations 
were performed to test the different simulation methods studied 
during the development of our method and to determine the 
optimal choices of parameters such as Courant factor and 
grid resolution. Full details of our method and complete results
will appear in Swesty \etal (1998).

\section{Results}

The goal of this study was to determine the effects of the choices 
of gravity coupling and reference frame
on the dynamics of the simulations.
The simulations that are described
as co-rotating were performed in a reference frame initially co-rotating
with the stars, and the simulations described as fixed frame
were performed in a fixed inertial reference frame. 
We use as a measure
of these effects the binary separation, the conservation of energy
and angular momentum, and the gravitational waveform. 
All simulations began with the same initial conditions describing two
$1.4 M_\odot$, $\gamma = 2$ polytropes with an initial separation
of 4.0$R_{\star}$ with 
$R_{\star}$ = 9.56 km. 
This initial separation is
large enough that the simulations should maintain stable
orbits. These simulations made use of 
a Courant factor of 0.4 and had
129 grid points along each axis, which with a simulation
cube size of 63.5 km gave a resolution of 0.5 km per zone.

Figure 1 shows the locations of the star centers and the location
of the center of mass for six NSM simulations that began
from the same initial data. The locations
\begin{figure}[]
\centering
\epsfig{file=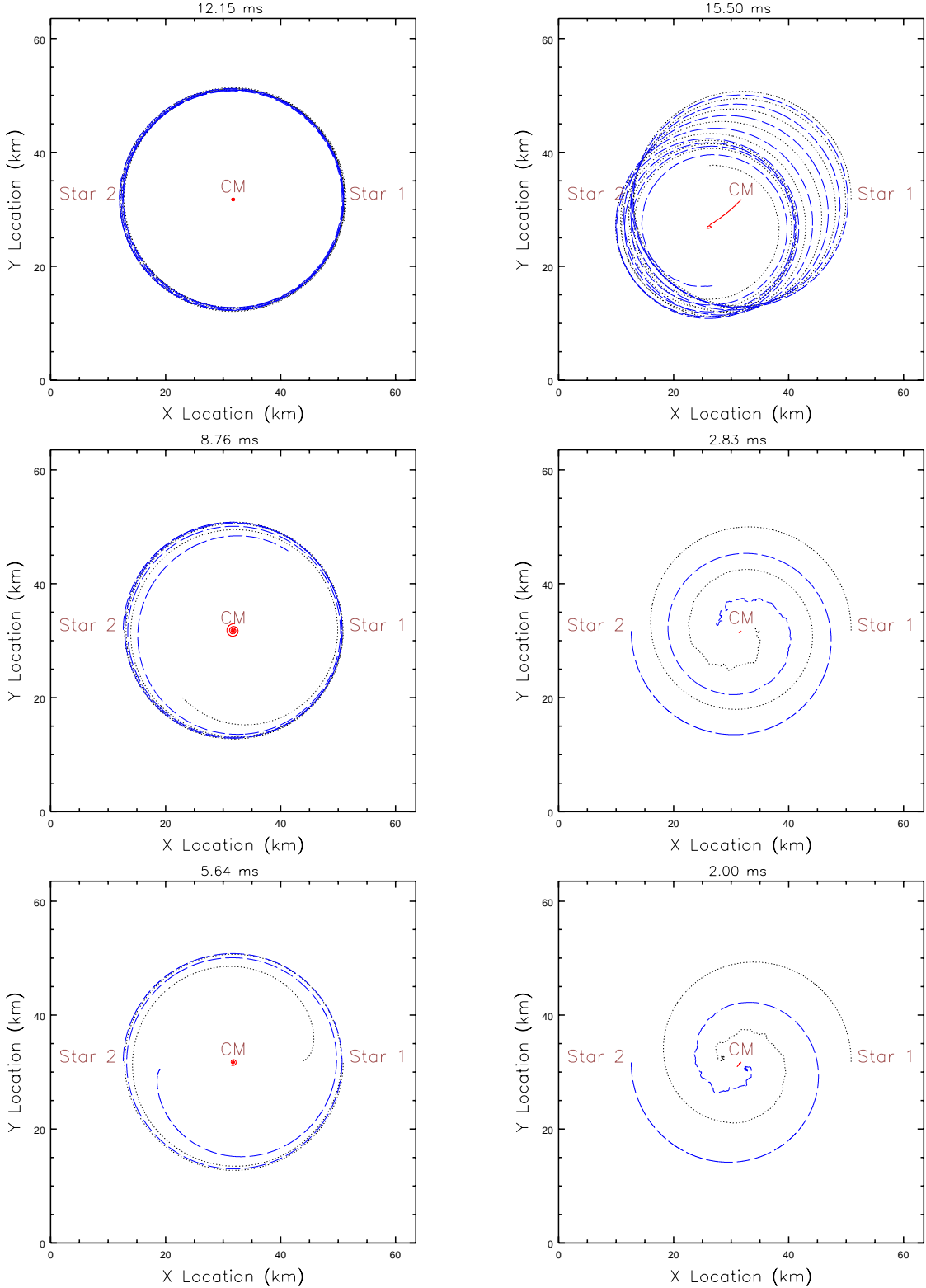,width=6.0in,angle=0}
\caption[3ff]
{Star locations and centers of mass from six NSM simulations
from the same initial conditions.
From top to bottom 
are results from simulations with time advanced, time centered,
and time lagged gravity. The images in the left column are 
from co-rotating simulations, and those in the right column are 
from fixed frame simulations.} 
\end{figure}
of the stars initially on the left-hand side of each frame
are shown as a dashed line, and the locations of the stars
initially on the right-hand side of each frame are shown as
a dotted line.
The three left-hand side panels are of simulations performed
in the co-rotating reference frame, 
and the right hand side panels are of simulations performed
in the fixed reference frame. The locations of the co-rotating
simulations have been mapped back into a fixed reference frame
for comparison.
In the figure, the instability of some of the simulations appears
as inward spiral deviations from perfectly circular orbits. 
The three simulations performed in the fixed frame 
(right-hand side panels) all demonstrate significant orbital 
instabilities. The three simulations performed in the
co-rotating frame were all more 
stable than their fixed frame counter parts.  
Simulations with time lagged gravity were the most unstable
for both fixed and co-rotating simulations. The most stable
simulation (the upper left panel) was performed in the 
co-rotating frame with time advanced gravity. It shows
very little deviation after 5 orbits (12.5 ms). The least stable
(the lower right panel) was performed in the fixed frame
with time lagged gravity. In it, the stars merged in
little more than one orbit (2.0 ms).
\begin{figure}[h]
\centering
\epsfig{file=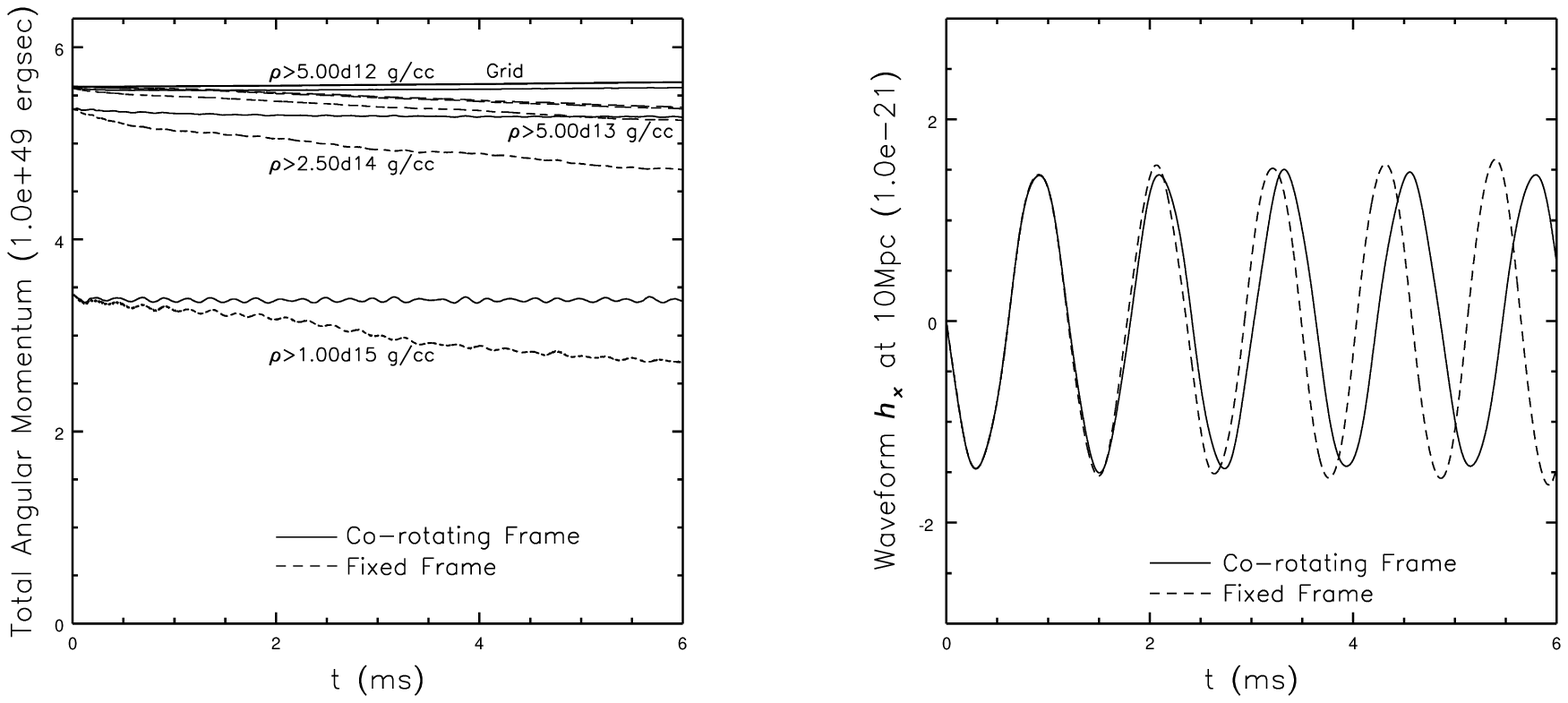,width=6.2in,angle=0}
\caption[3ff]
{Results of two time advanced gravity simulations for 
6.0 ms from the same initial conditions.  Plotted in the left 
panel is the angular momentum 
at five density thresholds as functions of time for a simulation in 
a fixed frame (dashed lines) and a simulation in an 
initially co-rotating frame (solid lines).
The right panel shows the $h_{\times}$ waveform for the fixed
frame simulation (dashed line) and the 
initially co-rotating frame simulation (solid line).}
\end{figure}

Figure 2 shows a comparison between the first 6.0 ms
of simulations with time advanced gravity in a
co-rotating frame (solid line) and in a fixed frame (dashed
line). The plots
are from the same simulations as the time advanced
simulations in Figure 1.
The panel on the left of Figure 2 shows the angular 
momentum at five density thresholds. Each line
represents the angular momentum of material with
density greater than a particular threshold density.  
Conserved angular
momentum would appear as a set of straight horizontal lines.
The plot shows that the simulation in the fixed frame does not 
conserve angular momentum as well as the simulation
in the co-rotating frame. Comparison of energy conservation
between the two simulations shows a similar result.
The panel on the right shows the $h_\times$ waveform
of the two simulations. While the waveforms have the
same amplitude, it is readily apparent that the two
become out of phase by approximately 1/4 orbit within 
less than three orbits. This significant difference 
shows that the choice of reference frame can have a strong
impact on the accuracy of a NSM simulation.  

\section{Conclusions}

Our results show that the simulation dynamics of NSMs
are very sensitive to the finite difference scheme with which
gravity is coupled to the hydrodynamics. We find that
simulations with a time advanced matter-gravity coupling 
are most stable in both co-rotating and fixed frame simulations. 
The co-rotating frame simulations are not as sensitive, however, to 
the coupling of gravity to the hydrodynamics.
Conservation of energy and angular momentum are 
important tests, and all future numerical calculations
of NSMs
should reveal how well these quantities are conserved. 
We find that simulations in a reference frame
initially co-rotating with the stars conserve energy and
angular momentum well.

Accurate waveform calculations are imperative for gravitational
wave astronomy, and consequently the waveforms must be free
from numerical artifacts.
Our results show that the waveforms of simulations in
a fixed frame can be very different from simulations with
the same initial conditions but performed in a co-rotating frame.
We find that waveforms become almost 1/4 orbit out of
phase within 6 ms in the most stable case of time
advanced gravity. The principle difference between
fixed and co-rotating frame simulations is that fixed frame simulations
dissipate angular momentum much more quickly than co-rotating
frame simulations. It is the requisite numerical advection 
and the need to resolve a rapidly varying gravitational
field in a
fixed frame that dissipates
angular momentum leading to the very different waveforms.

Based on these results, we conclude that NSM simulations should
be performed in a co-rotating reference frame. Numerical effects seem
to dominate the evolution of simulations in a fixed reference frame.
Further work is in progress to compare these calculations to those
done with other hydrodynamic algorithms such as the 
piecewise parabolic method (Colella and Woodward 1984).

\section*{Acknowledgments}
We would like to thank Bruce Fryxell, Mike Norman, 
and other members of our NSM Grand Challenge collaboration
for many helpful conversations on this material.
Computational resources were provided by NCSA and PSC
under Metacenter allocation
\#MCA975011. Funding for this research was provided by NASA
under NASA ESS/HPCC contract NCCS5-153.


\vspace{-14pt}
\begin{references}

\item[] Abramovici A, Althouse W E, Drever R W P,
Gursel Y, Kawamura S, Raab F J, Shoemaker D,
Sievers L, Spero R E and Thorne K S 1992 {\em Science} {\bf 256} 325
\item[] Colella P and Woodward P R 1984 {\em J. Comp. Phys.} {\bf 54} 174
\item[] Finn L S and Evans C 1990 {\em Astrophys. J.} {\bf 351} 588
\item[] Lattimer J M, Mackie F, Ravenhall D G and Schramm D N 1977
{\em Ap. J.} {\bf 213} 225
\item[] Lattimer J M, Pethick C G, Ravenhall D G and
Lamb D Q 1985 {\em Nucl. Phys.} A {\bf 432} 646
\item[] Lattimer J M and Swesty F D 1991 {\em Nucl. Phys.} A {\bf 535} 331
\item[] Norman M L 1980 Ph.D. thesis, Univ. California Davis,
        LLNL report UCRL-52946
\item[] Quashnock J M 1996 {\em Ap. J.} {\bf 461} L69
\item[] Rasio F A and Shaprio S L 1992 {\em Ap. J.} {\bf 401} 226
\item[] Rees M 1997 {\it Proceedings of the 18th Texas Conference
on Relativistic Astrophysics}
\item[] Stone J M and Norman M L 1992 {\em Ap. J. S.} {\bf 80} 791
\item[] Swesty F D, Wang E Y M and Calder A C 1998 in prep.
\item[] Thorne, K S 1995 {\it Proceedings of the Snowmass 95 Summer 
Study on Particle
and Nuclear Astrophysics and Cosmology}, eds. E W Kolb and R Peccei
\item[] Wang E Y M, Swesty F D and Calder A C 1998 this volume
\item[] Woods E and Loeb A 1994 {\bf Ap. J.} {\bf 425} L63


\end{references}
\end{document}